\documentclass[10pt,letterpaper]{article}
\usepackage{opex3}

\begin{document}

\title{Nonclassical 2-photon interference with separate intrinsically narrowband fibre sources}

\author{M.~Halder$^1$, J.~Fulconis$^1$, B.~Cemlyn$^1$, A.~Clark$^1$, C.~Xiong$^2$   , W.~J.~Wadsworth$^2$ and J.~G.~Rarity$^1$}

\address{$^1$Centre for Communications Research, Department of Electrical and Electronic Engineering, University of Bristol, Queen's Building, University Walk, Bristol, BS8 1TR, UK}
\address{$^2$Centre for Photonics and Photonic Materials, Department of Physics, University of Bath, Claverton Down, Bath, BA2 7AY, UK}
\email{matthaeus.halder@bristol.ac.uk}

\begin{abstract}

In this paper, we demonstrate a source of photon pairs based on four-wave-mixing in photonic crystal fibres.
Careful engineering of the phase matching conditions in the fibres enables us to create photon pairs at 597\,nm and 860\,nm in an intrinsically factorable state showing no spectral correlations. This allows for heralding one photon in a pure state and hence renders narrow band filtering obsolete. The source is narrow band, bright and achieves an overall detection efficiency of up to 21\% per photon. For the first time, a Hong-Ou-Mandel interference with unfiltered photons from separate fibre sources is presented.
\end{abstract}



\section{Introduction}

Single photons are one of the basic building blocks of quantum
information technologies. Numerous experimental
implementations of quantum computation schemes~\cite{klm},
quantum communication protocols~\cite{qcomm} as well as
quantum metrology~\cite{qmetrology} are based on single photons as the physical realization of the qubit. In order to code the state of such flying
qubits, quantum interferences are commonly exploited, which requires
the photons to be in a pure state~\cite{uren}.

The creation of photon pairs is a crucial tool to many of the
aforementioned applications, which are based on entangled 2-photon state (e.g.
entanglement swapping) or require heralding the presence of a single photon by detecting
its partner~\cite{heraldedSoujaeff}. Different approaches have been followed to produce photon pairs,
e.g. quantum dot emission~\cite{shields} or atomic
ensembles~\cite{thompson}, but implementations are still quite
cumbersome and in an early development stage. Frequency
mixing processes in nonlinear optical media, however, are well understood,
easier to realize and widely used~\cite{firstPDC, firstFWM}.

Still, a common drawback of all these techniques is the fact
that photon pairs are  in general created in a mixed state due to
correlations in energy and momentum~\cite{garay-palmett}. Thus, in order to obtain a photon pair in a pure
state, narrow spectral and spatial filtering is required. This introduces additional losses which limits the overall efficiency $\mu$ of detecting a created photon.
In experiments employing multiple photons, the probability $P_n$ of a successful detection of all n photons scales as $P_n\sim \mu^n$. In practice, this means that $n$ is usually limited to the order of 4-6 before prohibitively low n-fold coincidences render a successful experiment impossible within a reasonable period of time.
So far, high detection efficiency on the one hand and narrow filtering on the other hand have been mutually exclusive. In this paper, we present a source of photon pairs based on 4-wave mixing (FWM) in photonic crystal fibres (PCF) featuring both a high overall detection efficiency and narrowband photons.

Due to the careful engineering of the phase matching, the pairs are created in an intrinsically factorable state, i.e. free of correlations in any degree of freedom. In our case this means, that signal and idler photon have decoupled spectral distributions without any frequency correlations between them. Hence, the detection of an idler photon heralds the presence of a signal photon in a pure state~\cite{grice}, which makes additional filters obsolete. As a consequence, losses are reduced and future multi-photon experiments with $n >4$ become possible.

\section{FWM in Photonic Crystal Fibres}

\begin{figure}[htbp]
\includegraphics[width=7cm]{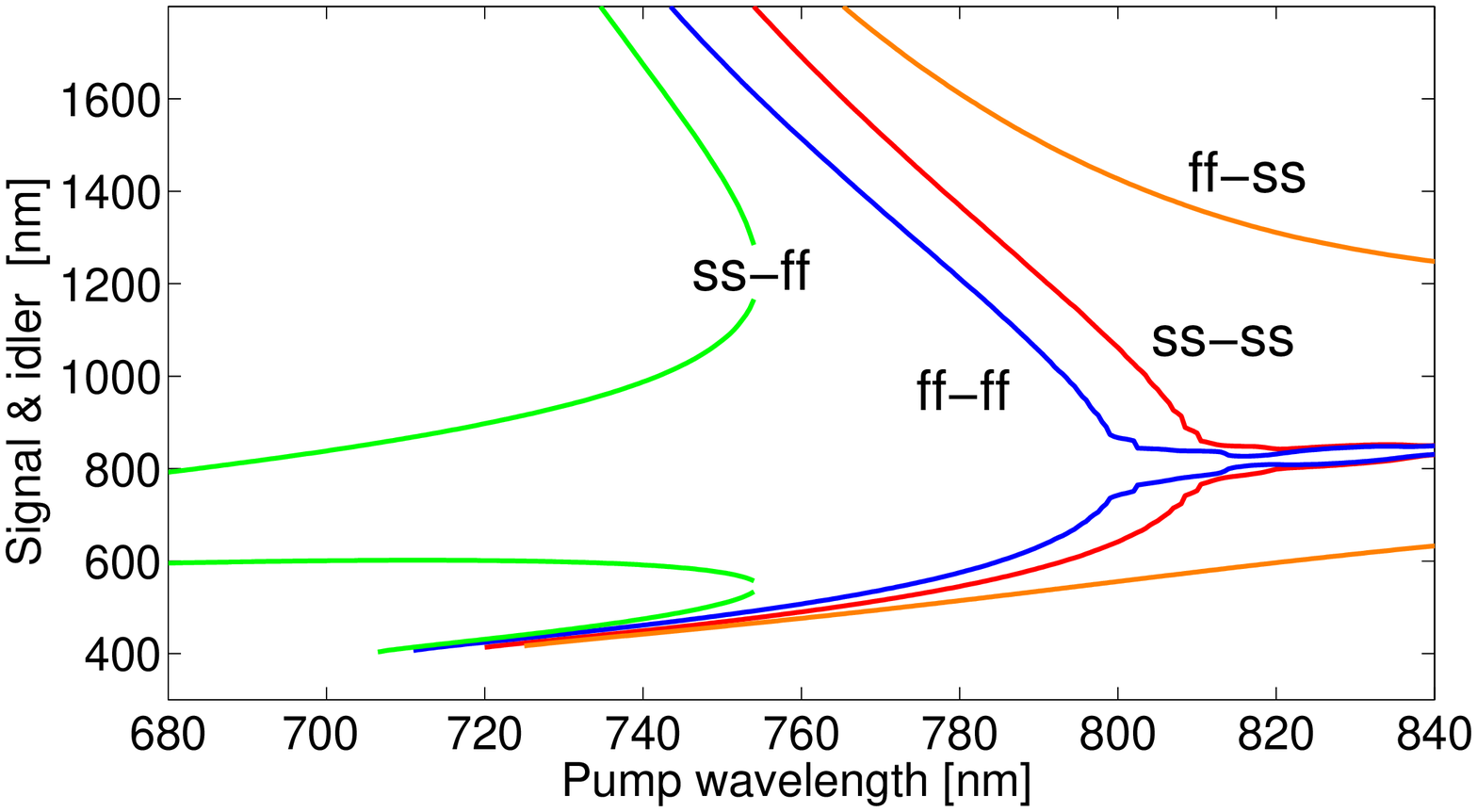}
\hspace{1cm}
\includegraphics[height=3.4cm]{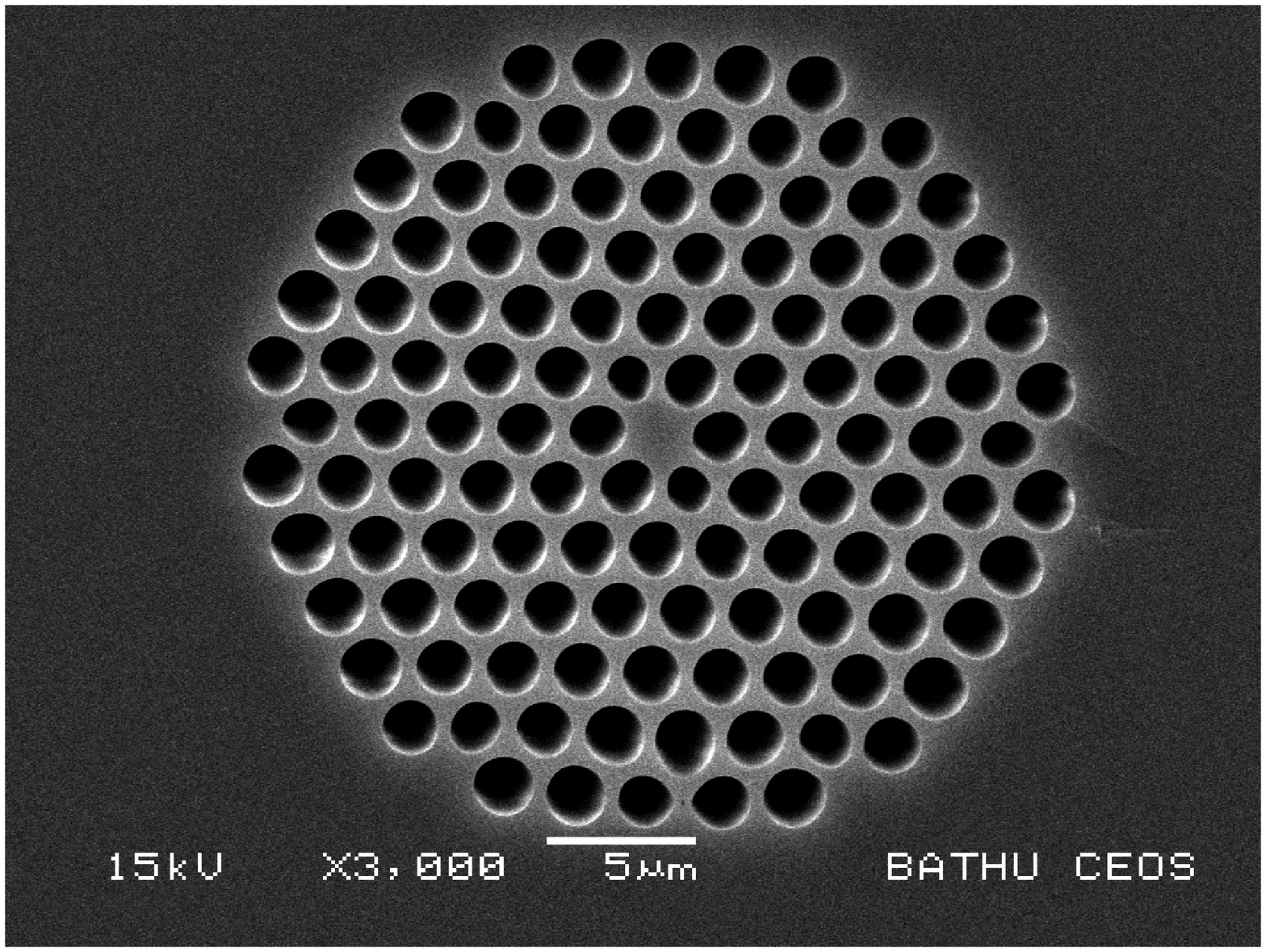}
\caption{(a) Phase matching curve for FWM in birefringent PCF.
The curves ff$\rightarrow$ff (ss$\rightarrow$ss) represent the cases where pump and daughter photons are all polarized along the f (s) axis. Note that the same curve is obtained for a non-birefringent fibre described by just one refractive index.
In the case ff$\rightarrow$ss (ss$\rightarrow$ff), orthogonally polarized pump and daughter photons see different modal indices which leads to a region in the phase matching curve featuring a horizontal tangent. At this point signal and idler photons have no frequency correlation and are hence in a factorable state. (b) SEM of the central region of the fibre used in these experiments. The birefringence required was introduced by reducing the size of two holes adjacent to the core (11 o'clock and 5 o'clock).}
\label{fig:phasematch}
\end{figure}

In FWM, two photons from an intense pump field are converted into two daughter photons, signal and idler. This process takes place according to conservation of energy and phase matching

\begin{equation}
2\omega_p = \omega_s + \omega_i  \hspace{1.3cm} \textrm{and} \hspace{1.3cm} 2 \frac{n_p \omega_p}{c} -  \frac{n_s \omega_s}{c} - \frac{n_i \omega_i}{c} - 2\gamma P = 0
\label{eq:pm}
\end{equation}

\noindent with $n_{p,s,i}$ the refractive index at pump, signal and idler frequency $\omega_{p,s,i}$, respectively, $\gamma$ the nonlinear coefficient of the fibre~\cite{agrawal} and $P$ the peak pump power.
FWM in PCF is a well understood process and has already been used for pair-photon generation. Both cases where the pump, signal and idler waves are all in the same polarization~\cite{friberg, fulconisPRL} as well as in orthogonal polarization states~\cite{fan} have been studied.
With an elliptical fibre core or an asymmetric cladding, we obtain a slight birefringence and different effective indices $n(\lambda)$ along the fast~($f$) and slow~($s$) axis of the fibre. Solving Eq.~(\ref{eq:pm}) results in the phase matching curves as plotted in Fig.~\ref{fig:phasematch}(a), where \textit{ss}$\rightarrow$\textit{ss} and \textit{ff}$\rightarrow$\textit{ff} correspond
to cases where all fields are polarized in the same direction ($n_{s,i}(\lambda)=n_p(\lambda)$) and \textit{ss}$\rightarrow$\textit{ff} and \textit{ff}$\rightarrow$\textit{ss} correspond to cases where the pump field is orthogonal to the signal and idler fields ($n_{s,i}(\lambda) = n_p(\lambda) \pm \delta n(\lambda)$).
If we ignore the wavelength dependence of the birefringence, $\delta n(\lambda)$, then to first order this appears in Eq.~(\ref{eq:pm}) in the same way as the term $2\gamma P$, but with either a positive or negative sign, and usually much greater magnitude.

We fabricated several fibres with zero dispersion wavelengths close to 800\,nm and different birefringence by controlling the hole sizes during the fibre draw~\cite{continuum}. The calculated phase matching curves for the fibre selected for these experiments are shown in Fig.~\ref{fig:phasematch}(a) and an electron micrograph of the fibre is shown in Fig.~\ref{fig:phasematch}(b).

For carefully controlled $\lambda_p$, $n_p$ and $n_i$, we can achieve a situation where the group velocity of the pump photons (e.g. in the $s$-axis) equals the group velocity of one of the daughter photons, the idler in our case (e.g. in the $f$-axis).
In this configuration, due to matching group velocities of pump and idler light, the idler photon is maximally confined in time and hence minimally in energy, and vice versa for the signal.
As a result, the entire energy uncertainty of the photon pair is concentrated on the idler photon and the temporal uncertainty on the signal photon. Furthermore, over a limited range the signal wavelength now becomes independent of the pump wavelength, which translates into a horizontal tangent on the phase matching curve, as can be seen in Fig.~\ref{fig:phasematch}(a) for $\lambda_p$=705\,nm. This intrinsic lack of energy correlation between the two daughter photons means that their joint spectral amplitude (JSA) can be written in the factorable state \mbox{$F(\omega_s,\omega_i)=S(\omega_s)\times I(\omega_i)$}. See also~\cite{garay-palmett} for more details.

By thoroughly controlling the absolute and relative index of the two fibre axes via the core diameter and size of the air holes in the silica cladding, we are able to tailor the PCF to make it perfectly suited to our application. This enables us to build a photon pair source with high single mode coupling efficiencies
because no lossy filtering is required anymore and the photons, created and propagating in the fundamental spatial mode of the PCF, can efficiently be coupled into any other single mode fibre.

\begin{figure}[h]
\includegraphics[width=0.31\columnwidth, height=4cm]{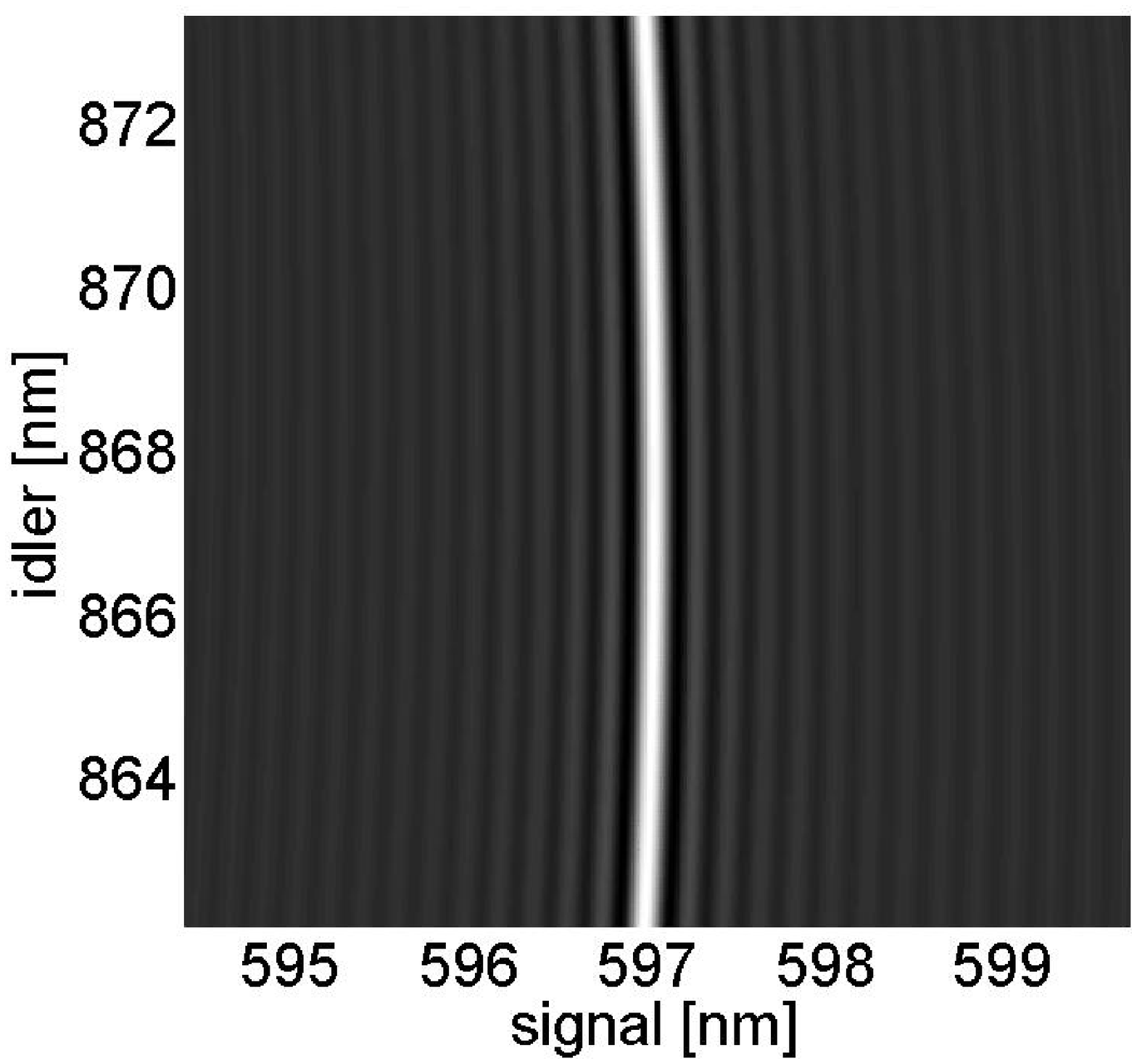}
\includegraphics[width=0.018\columnwidth]{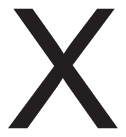}
\includegraphics[width=0.31\columnwidth,height=4cm]{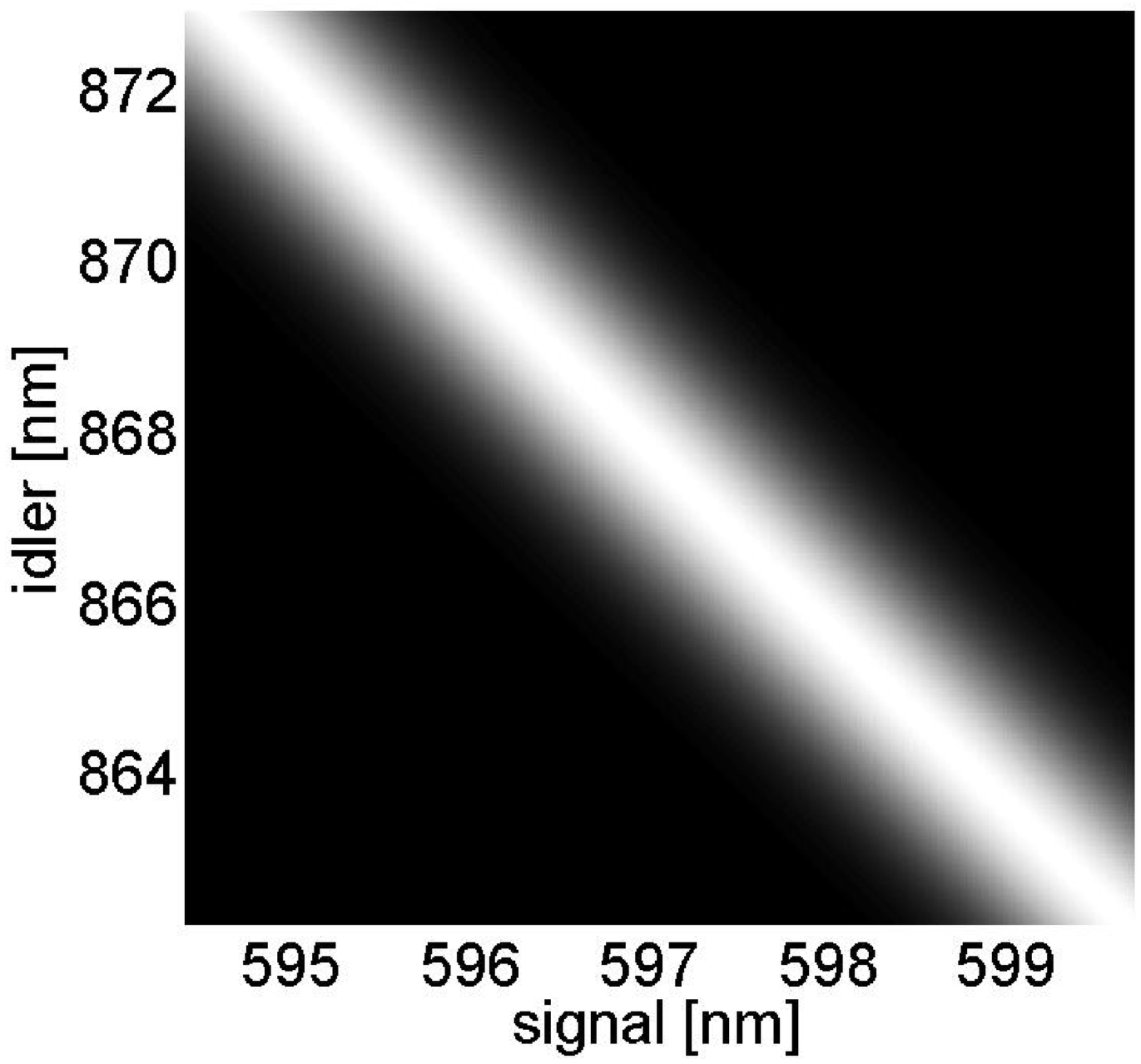}
\includegraphics[width=0.019\columnwidth]{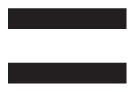}
\includegraphics[width=0.31\columnwidth,height=4cm]{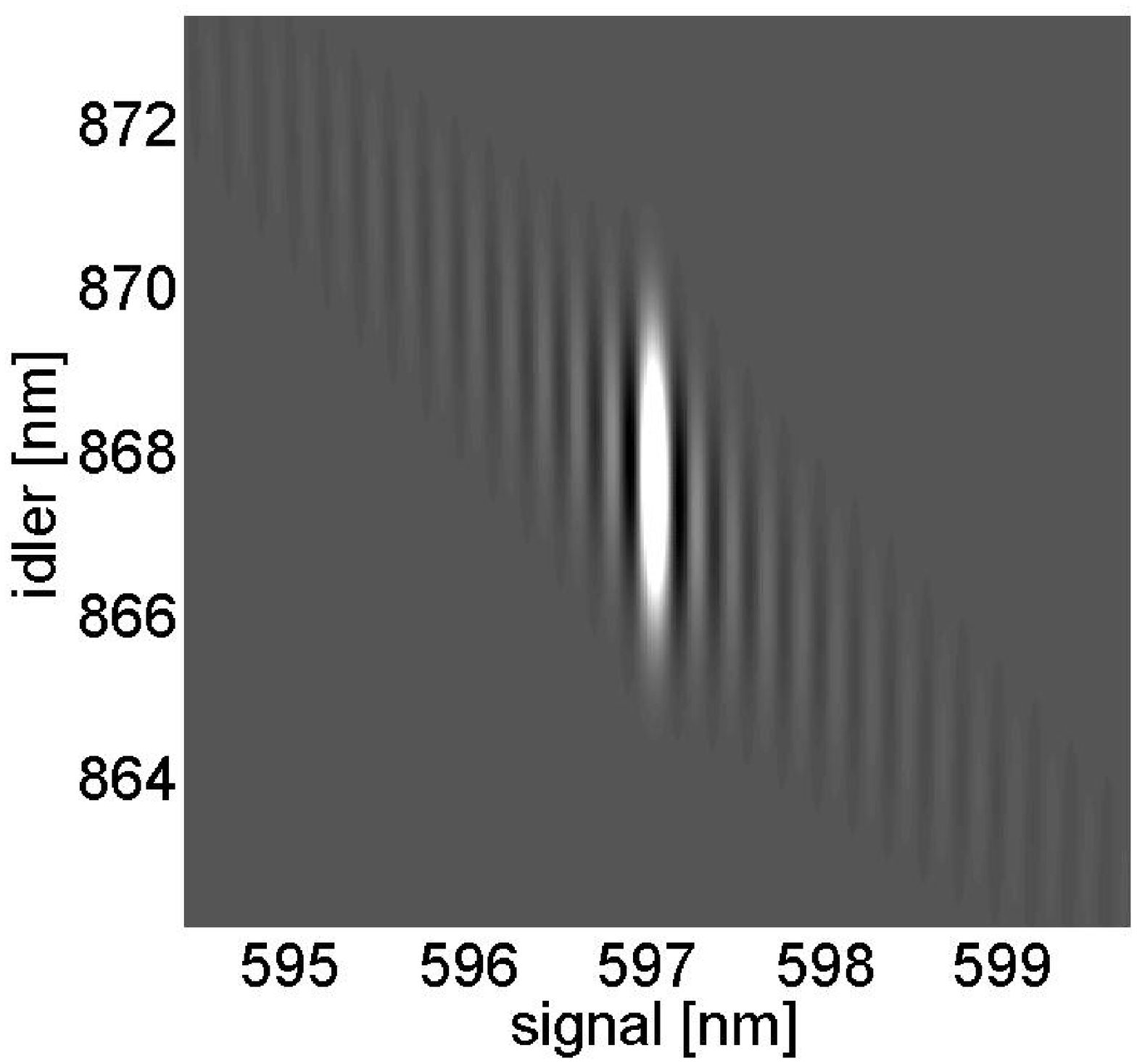}
\caption {(a)~Phase matching condition of the PCF, (b)~Envelope of the pump intensity, (c)~JSA of the created photon pair.}
\label{fig:3}
\end{figure}

Figure~\ref{fig:3}(c) shows the JSA as a function of signal and idler wavelength, given by the product $\phi \times \alpha$ of the natural phase matching condition of the PCF $\phi = sinc(\Delta kL/2)$ (Fig.~\ref{fig:3}(a))
and the pump envelope function $\alpha = exp(-(\Delta\omega_s+\Delta\omega_i)^2/8\sigma^2)$ (Fig.~\ref{fig:3}(b)) with $\Delta k $ and $\Delta\omega_{s,i}$ the detuning from the perfect phase matching $k_{s,i}$ and $\omega_{s,i}$, respectively, and $L$ the length of the fibre  \cite{alibart06}. In this picture, the JSA of an ideal factorable state with a Gaussian spectral distribution is represented by an ellipse with symmetry along the horizontal and vertical axes.
However, in the real case (Fig.~\ref{fig:3}(c)~and~\ref{fig:ripples}) we can see that although the central peak may be well approximated by an elliptical Gaussian there are significant outlying peaks which can make the state partly correlated.

\section{The experimental setup}

\begin{figure}[htbp]
\centering\includegraphics[width=7cm]{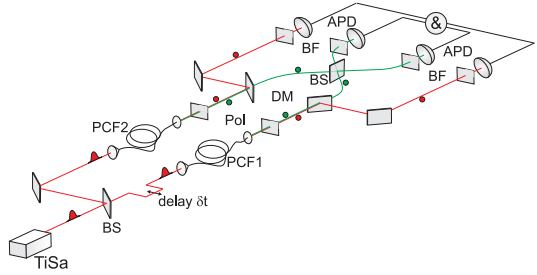}
\caption{Schematic of the experimental setup. See text for explanations.}
\label{fig:setup}
\end{figure}

In order to show the suitability of our source for quantum communication or quantum gates~\cite{takeuchiCnot}, we will now demonstrate non-classical 2-photon interference. Recently, such quantum interferences have been reported for photons created in a single source~\cite{walmsleyDip}. Here, we show non-classical interference between photons originating from two separate sources, as required for example for long distance quantum relays or quantum repeaters~\cite{qcomm}. The experimental setup is shown in Fig.~\ref{fig:setup}.

\begin{figure}[htbp]
\centering
\includegraphics[width=0.35\columnwidth]{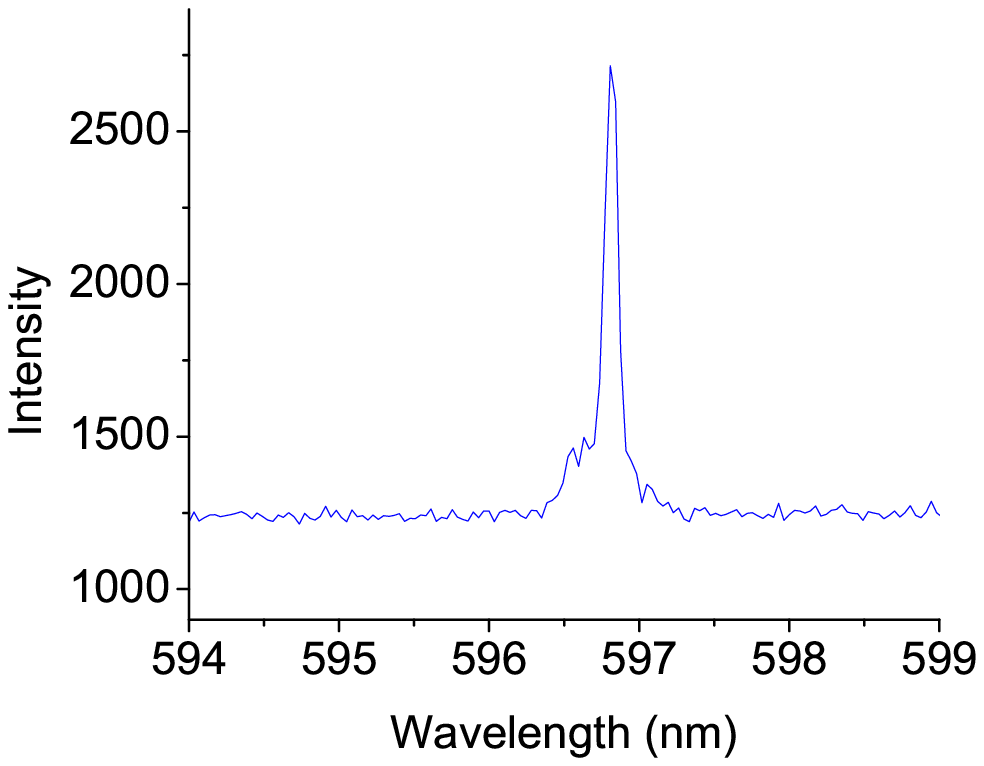}
\includegraphics[width=0.28\columnwidth]{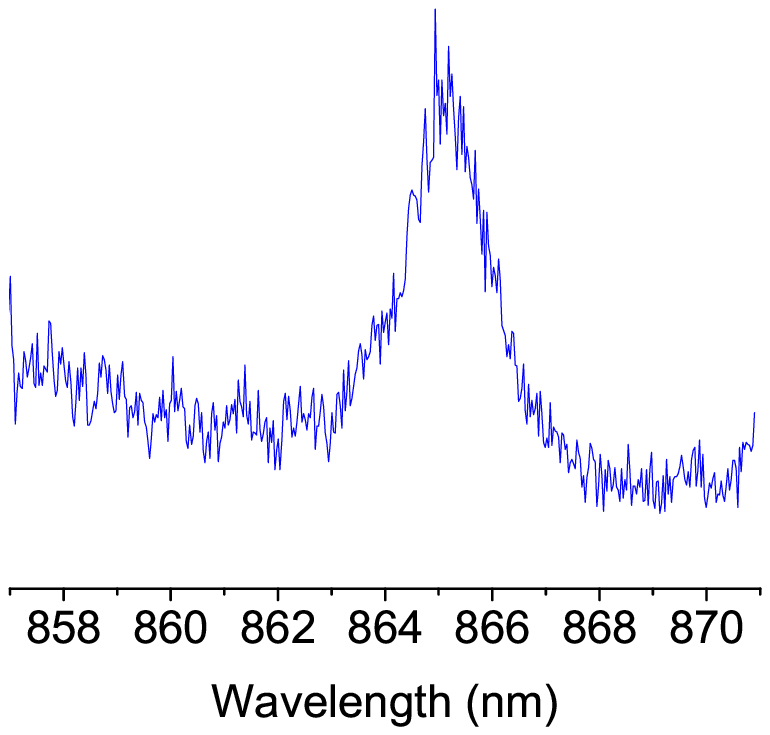}
\caption{Intrinsic spectral distribution of created signal and idler photons, measured by a spectrometer and a cooled CCD camera.}
\label{fig:signalidler}
\end{figure}

Two separate 40\,cm long PCFs (PCF1, PCF2) are pumped by a mode-locked Ti:Sapphire laser in fs-mode at 705\,nm. This gives rise to pairs of signal and idler photons at 597\,nm and 860\,nm respectively. The bandwidth $\Delta\lambda_p$ of the pump laser is filtered to 0.9\,nm FWHM, resulting in a coherence time of $\tau_p$=0.8\,ps. The intrinsic bandwidth of the signal photon created is as narrow as 0.13\,nm corresponding to a coherence time $\tau_s$=4ps\,$>\tau_p$ and the idler bandwidth is 2\,nm. Fig.~\ref{fig:signalidler} shows the spectra of the created photons, taken with a spectrometer and a liquid nitrogen-cooled CCD camera, integrated over 1\,s. 

At the output of each PCF, a polarizer in parallel orientation with respect to that of the created photons blocks the remaining pump field and a dichroic mirror (DM) separates the daughter photons in frequency. The idler photons are passed through broadband interference filters (BF, FWHM 10\,nm$\gg\Delta\lambda_i $) to reject spontaneous Raman emission and residual pump light, and are then directed onto single-photon Si-avalanche photon detectors (APDs). The successful detection of an idler photon heralds the presence of a signal photon in the corresponding source. Such a 2-photon coincidence measurement is shown in \cite{alibart06}. The heralded signal photons are then sent into the two input ports of a fibre 50/50 beam splitter (BS), passed through broad band filters (BF, FWHM~40\,nm$\gg\Delta\lambda_s$) and detected by APDs, one at each output port.

The overall collection efficiency of our signal and idler photons are 0.21 and 0.18, respectively. This includes all losses due to absorption in the PCF, transmission of the blocking filters (81\% and 65\%, respectively) and dichroic mirror, coupling to single-mode fibres, and detector efficiencies (59\% and 40\%, respectively).

\section{Results and discussions}

\begin{figure}[htbp]
\centering\includegraphics[width=7cm,height=4cm]{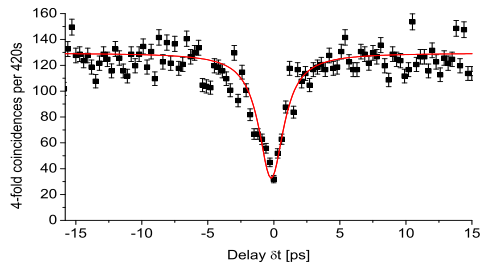}
\caption{4-fold coincidence count rate as a function of the temporal delay $\delta t$ between the signal photons impinging on the beam splitter. The error bars are determined for a Poisson distribution, given by the square root of the obtained count rates. The solid curves is a fit through the experimental data with a Lorentzian function resulting in a visibility of 76.2\%.}
\label{fig:dip}
\end{figure}

To demonstrate the intrinsic purity of the heralded photons we show a Hong-Ou-Mandel interference \cite{HOM87} between two signal photons originating from different sources and sent into different input ports of a 50/50 BS. For photons, indistinguishable in spatial mode, polarization, wavelength and arrival time, photon bunching occurs and leads to detection of the two photons always in the same output port of the BS. Hence the coincidence detection rate between the two output ports will drop to zero.
To ensure the modal indistinguishability of the two signal photons, we use a single mode fibre BS for perfect spatial overlap. The signal polarization in input port~1 is manually controlled in order to match the polarization mode in port~2. Spectral overlap of the two signal photons is obtained by temperature controlling the PCFs independently by a Peltier element in order to tune their phase matching condition and hence the wavelength of the emitted signal photons. The temperature dependence of the central signal wavelength is 11\,pm/$^{\circ}$C, making it relatively easy to control as a stability of $\pm1\,^{\circ}$C suffices. By varying the relative arrival time of two heralded signal photons at the BS ($\delta t$) and recording the 4-fold coincidence probability of detecting each photon in a separate APD, a Hong-Ou-Mandel dip can be demonstrated. The experimental result is shown in Fig.~\ref{fig:dip}. Note that the coincidence count rate for two heralded signal photons, one in each output port, is constant for $\delta$t$\neq$0, but dropping for photons arriving simultaneously at the BS ($\delta$t=0).

\begin{figure}[htbp]
\centering\includegraphics[width=10cm,height=4.8cm]{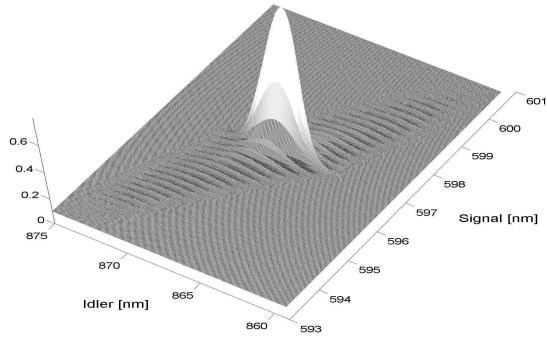}
\caption{JSA of the bi-photon state as a function of signal and idler wavelength. It can clearly be seen that outside the central peak, oscillations given by the sinc-shaped phase matching introduce some asymmetry, leading to distinguishability and hence lack of visibility.}
\label{fig:ripples}
\end{figure}

Fitting the uncorrected experimental data with a Lorentzian function, a raw visibility for the non-classical interference of $76.2\pm2.1\%$ is observed. The limited visibility can be attributed to different reasons: the non-perfect splitting ratio of the BS (1\%), noise due to creation of multiple pairs ($<$1\%) and Raman emission ($<$1\%), and difference in the spectra of the two signal photons leading to distinguishability. However the main reason is frequency-correlation in the JSA of a signal-idler pair ($\sim$20\%).
Looking at Fig.~\ref{fig:3}(c) and \ref{fig:ripples} we see that the sinc function nature of the phase matching function -given by the Fourier transform of the temporal (top hat) creation window of a pair in the PCF- leaves significant ripples along the energy matching direction. Hence in a small way signal and idler photons are correlated. Detection of an idler photon gives partial information on the signal photon. This reduces the indistinguishability, and hence the interference visibility, of the heralded signal photons. Using the theory derived in \cite{mosley} we have simulated our PCF and find good agreement with our experimental results \cite{BenLongPaper}.

This slight distinguishability increases the Schmidt number $K$, a measure of the frequency correlations between signal and idler photons (see~\cite{mosley}). For a factorable state $K=1$ and for any mixed state $K>1$.
It can be shown that the visibility will approach 100\% (i.e. $K\rightarrow 1$) for increasing fibre lengths. However, this only holds true for a theoretical PCF, homogeneous over an infinite length $L$. Due to fluctuations during the fabrication process of our PCF, we are restricted to $L\approx$1\,m, before phase matching conditions vary too much.

Another way to reduce distinguishability, and hence increase visibility, is the use of a PCF featuring a higher birefringence $\delta n(\lambda)$. Phase matching conditions in such a fibre result in a wider spectral separation of signal and idler photons. Consequently, detection efficiencies of the idler photons will decreased due to low sensitivity of the Si-APD beyond 900\,nm, leading to a reduced overall coincidence count rate.

\section{Conclusion}

We have demonstrated the creation of photon pairs via four wave mixing in photonic crystal fibres. The pairs are created in an intrinsically factorable state without any spectral filtering.
This allows us to herald the presence of a photon in a pure state and to show Hong-Ou-Mandel interference
between photons originating from separate sources with a raw visibility of 76.2\%. Furthermore we achieve an overall detection efficiency of up to 21\%, making future multi-photon experiments feasible due to high multi-fold coincidence count rates.
For instance in a 6-photon setup with 80\,MHz repetition rate, 20\% overall detection efficiency and 0.1 photons per pulse we will achieve a 6-fold coincidence count rate of 5.1 per second, sufficient for an experiment with a reasonably short integration time.

\section*{Acknowledgments}

The work was supported by the UK EPSRC (QIP IRC and EP/F002424/1), the EU Integrated Project Qubit Applications (IP QAP) and ACDET. W.J.W. is a Royal Society University Research Fellow, J.G.R. is supported by a Wolfson Merit award and M.H. is supported by an AXA fellowship.

\end{document}